\documentclass{aa}

\usepackage{epsf}
\usepackage{txfonts}
\usepackage{color}

\def\changed{}
\def \etal   {\hbox{et~al.\/}}
\def\nchange{}

\newcommand{\mdot}{\mbox{$\dot{M}$}}
\newcommand{\lsim}{\raisebox{-.4ex}{$\stackrel{<}{\scriptstyle \sim}$}}
\newcommand{\msim}{\raisebox{-.4ex}{$\stackrel{>}{\scriptstyle \sim}$}}

\begin{document}

\title{Neglecting the porosity of hot-star winds
can lead to underestimating  mass-loss rates}

\author{L. M. Oskinova \and W.-R. Hamann \and A. Feldmeier} 

\offprints{L. Oskinova \\
\email{lida@astro.physik.uni-potsdam.de}}

\institute{Lehrstuhl Astrophysik der Universit\"at Potsdam,
Am Neuen Palais 10, D-14469 Potsdam, Germany}

\date{Received 18 April 2007 / Accepted 18 September 2007}

\abstract
{The mass-loss rate is a key parameter of massive stars. Adequate
stellar atmosphere models are required for spectral analyses and
mass-loss determinations. Present models can only account for the
inhomogeneity of stellar winds in the approximation of small-scale
structures that are optically thin. Compared to previous homogeneous
models, this treatment of ``microclumping'' has led to reducing 
empirical mass-loss rates by factors of two to three. Further
reductions are presently discussed in the literature, with far-reaching
consequences e.g.\ for stellar evolution and stellar yields.} 
{Stellar wind clumps can be optically thick in spectral lines. 
We investigate how this ``macroclumping'' influences the 
radiative transfer and the emergent line spectra and discuss its 
impact on empirical mass-loss rates.} 
{The Potsdam Wolf-Rayet (PoWR) model atmosphere code is generalized in
the ``formal integral'' to account for clumps that are not necessarily
optically thin. The stellar wind is characterized by the filling
factor of the dense clumps and by their average separation.  An
effective opacity is obtained by adopting a statistical distribution of
clumps and applied in the radiative transfer.}
{{Optically thick clumps reduce the effective opacity. This has a
pronounced effect on the emergent spectrum. Our modeling for the
O-type supergiant $\zeta$\,Puppis reveals that the optically thin
H$\alpha$ line is not affected by wind porosity, but that the P\,{\sc
v} resonance doublet becomes significantly weaker when macroclumping
is taken into account. The reported discrepancies between
resonance-line and recombination-line diagnostics can be resolved
entirely with the macroclumping modeling without downward revision of
the mass-loss rate.  In the case of Wolf-Rayet stars, we demonstrate
for two representative models that stronger lines are typically
reduced by a factor of two in intensity, while weak lines remain
unchanged by porosity effects.}}
{{\nchange Mass-loss rates inferred from optically thin emission, such
as the H$\alpha$ line in O stars, are not influenced by macroclumping.
The strength of optically thick lines, however, is reduced because of  
the porosity effects. Therefore, neglecting the porosity in stellar
wind modeling can lead to underestimating empirical mass-loss rates.}}


\maketitle

\section{Introduction}

Mass loss by stellar winds plays a key role in the evolution of
massive stars. The feedback of chemically enriched material, momentum,
and mechanical energy is important for stellar clusters and the
formation of stars, so it has consequences for the whole
cosmic evolution {\changed (e.g. Leitherer et al.\ \cite{lei92};
Woosley et al.\ \cite{woo93})}.

It is very important to know the mass-loss rate
of stars in their different evolutionary phases, and its dependence on
initial mass, composition, and possibly further parameters. Decades of
efforts have been needed to establish reliable mass-loss rates for OB
and Wolf-Rayet (WR) type stars {\changed (see Kudritzki \& Puls
\cite{kp00})}.  Nevertheless, there are doubts whether the presently
accepted mass-loss rates {\nchange are correct}. Some recent
papers favor a dramatic {\nchange downwards reduction} of O-star
mass-loss rates with far-reaching consequences.  {\nchange The reason
for the uncertainties largely stem from the implications of wind
inhomogeneity, i.e.\ ``clumping''.}

{One of the numerous proofs of wind clumping is} the line profile
variability routinely found in O and WR spectra, where different
types of variability can be distinguished. ``Discrete absorption
components'' (DACs) are observed regularly in the UV spectra from O
stars {\changed (e.g. Kaper et al.\ \cite{kap96})}. They display
periodic patterns on timescales of days and are most likely
associated with stellar rotation, but their slow wavelength drift is
still enigmatic (e.g.\ Hamann et al.\
\cite{DAC-paper}). ``Modulations'', which have a slightly shorter
timescale and are not strictly periodic, are thought to reflect
non-radial stellar pulsations (Eversberg et al.\
\cite{Eversberg_etal1998}). Only the stochastic variability on short
timescales of hours is attributed to wind clumping. Such stochastic
variability has been studied by L\'epine \& Moffat
(\cite{Lepine+Moffat1999}) in WR stars and by Markova et al.\
(\cite{Markova_etal2005}) in the H$\alpha$ line of O stars.

Clear but indirect evidence for hydrodynamic shocks in O-star winds
is the observed X-ray emission and the superionization that the
X-rays produce in the cool plasma (Cassinelli \& Olson \cite{cas79}).
Theoretical work has attributed such shocks to the intrinsic
``de-shadowing'' instability of line-radiation driven winds (Owocki et
al.\ \cite{OCR1988}; Feldmeier \cite{Feldmeier1995}).

{A first approximation to account for clumping in stellar wind models has
been introduced about a decade ago  (Hamann \& Koesterke
\cite{HK98-clumping}; Hillier \& Millier \cite{HiMi99}; Puls \etal\
\cite{Puls_etal2006}). It is based on the assumption that the wind
clumps are small compared to the mean free path of the photons. This
``microclumping'' approximation introduces a new parameter $D$ to
describe the factor by which the density in the clumps is enhanced
compared to a homogeneous model with the same mass-loss rate. The most
important consequence of accounting for wind clumping within the
optically-thin clump limit is the reduction of the empirical mass-loss
rates by a factor $\sqrt{D}$. This holds in principle for all
diagnostics that depend quadratically on the density, thus
including all emission lines in WR spectra, thermal radio emission, and
the H$\alpha$ line from O stars. However, the different diagnostics
may be affected to a different degree, due to the possible radial
variation in the clumping factor $D$.

A handle for estimating the clumping factor in Wolf-Rayet atmospheres
is provided by strong emission lines with extended line wings,} caused
by the frequency redistribution of line photons scattered by free
electrons. Clumping reduces the relative strength of these wings,
compared to homogeneous models (Hillier \cite{Hillier1991}), because
typical WR emission lines scale with the square of density, while
electron scattering scales linearly.  Detailed modeling of electron
scattering wings therefore allows the density enhancement factor $D$
to be estimated.

Hamann \& Koesterke (\cite{HK98-clumping})  typically found
$D$=4 for a few selected Galactic WN stars, corresponding to a volume
filling factor of the clumps of $f_{\rm V}$ = 0.25. More recent
studies (Crowther et al.\ \cite{Crowther_etal2002}) prefer even
stronger clumping, with filling factors smaller than $f_{\rm V}$ = 0.1.

For O-stars, Puls et al.\ (\cite{Puls_etal2006}) conclude that the
clumping factor $D$ is about four times larger in the line-forming
region, compared to the radio-emitting region far away from the
star. Resonance lines, which are {\nchange not affected by}
microclumping because of their linear density-dependence, are mostly
saturated in O stars and therefore not suitable for precise mass-loss
determinations.  {\nchange However, Fullerton et al.\
(\cite{Fullerton06}) exploit the fact that the far-UV spectral range
observed with FUSE contains the P\,{\sc v} resonance line, which is
typically unsaturated in O-star spectra, and analyze this line with
the SEI method (\cite{sei}).}  They derived mass-loss rates that are
about ten times lower than obtained from $\rho^2$ diagnostics with
unclumped models. Very strong clumping and consequently low mass-loss
rates were also obtained by Bouret et al.\ (\cite{Bouret_etal2005})
for two O stars analyzed with non-LTE model atmospheres. Such low
mass-loss rates would have dramatic consequences on stellar evolution
and feedback.

However, all these downward-revisions of mass-loss rates are based on
the assumption that the clumps are always smaller than the mean free
path of photons. For strong lines in dense winds, this approximation
cannot be justified. In the present paper we relax this small-clump
approximation and account for clumps of arbitrary optical depth.
Significant work has already been done to unveil the effects of
optically thick inhomogeneities in stellar atmospheres.

{\em Grey opacity} in a stellar atmosphere where the photon mean free
path does not exceed the scale of wind inhomogeneities was considered
by Shaviv (\cite{sh98}). The term ``porous atmosphere'' was coined by
Shaviv (\cite{sh00}) to describe a multi-phase medium that allows more
radiation to escape {\nchange while exerting a weaker average force.}
Such an atmosphere yields a considerably lower mass-loss rate for the
same total luminosity and can explain the apparent existence of
super-Eddington stars (Shaviv \cite{sh00}).  Owocki \etal\
(\cite{ow04}) introduced a ``porosity-length'' formalism to derive a
simple scaling for the reduced effective opacity and used this to
obtain an associated scaling for the continuum-driven mass-loss rate
from stars that formally exceed the Eddington limit.

{\em Continuum opacity} in inhomogeneous winds with optically thick
clumps was studied by Feldmeier \etal\ (\cite{Feldmeier2003}). They
considered X-ray lines emitted by an optically thin hot plasma that is
attenuated in a fragmented cool stellar wind by strong continuum
opacity from bound-free and K-shell photoionization processes.
Oskinova \etal\ (\cite{Oskinova2004}) show that wind porosity can
explain {\nchange X-rays line profiles and the weak}
wavelength-dependence of the effective opacity. Considering the
specific case of spherical clumps, {\changed Owocki \& Cohen
(\cite{oc06}) argue that a large clump separation is required to fit
the observed X-ray emission line profiles. However, Oskinova \etal\
(\cite{Oskinova2006}) reproduced the observed emission line profiles
in X-ray spectra of O-stars with a wind model assuming plausible model
parameters. Addressing mass-loss rates inferred from the analysis of
radio measurements, Brown \etal\ (\cite{brown04}) point out that 
optically-thick clumping leads to the reduction of multiple scattering 
and, consequently, photon momentum delivery.}

{\em Line opacity} in inhomogeneous stellar winds is modeled for the
first time in the present paper. Considering the effect of clumping on
UV resonance lines, Massa \etal\ (\cite{massa03}) and Fullerton \etal\
(\cite{Fullerton06}) discuss how porosity can affect the formation of
P Cygni lines and, in extreme cases, produce an apparently unsaturated
profile for a line that would be extremely saturated if the wind
material were distributed smoothly.

In the following section (Sect.\,2) we briefly review the line
radiative transfer in the conventional optically-thin clump
approximation.  In Sect.\,3 we describe a statistical treatment for
clumps of arbitrary optical thickness. In Sect.\,4 both models are
compared with observed mass-loss diagnostic lines in the spectrum of
$\zeta$\,Puppis.  {\changed A brief parameter study is conducted in
Sect.\,5.}  Two representative examples of Wolf-Rayet model spectra
are presented in Sect.\,6, and a summary is given in Sect.\,7.

\section{{Radiative transfer in the limit of optically thin clumps}}

Radiative transfer in an inhomogeneous medium is implemented in
stellar-wind codes like CMFGEN and PoWR to a first approximation.
This approximation holds in the limit that the size of the structures
is small compared to the mean free path of photons (microclumping).
Moreover, it is assumed that the interclump medium is void.  The
density in the clumps is enhanced by a factor of $D$ compared to the
density {$\rho$} of a smooth model with the same mass-loss rate
{\mdot}; hence, the volume filling factor is $f_{\rm V} = D ^{-1}$,
where $D$ may depend on the location in the stellar wind, i.e.\ on the
radial coordinate $r$.

As there is no matter outside the clumps, rate equations are solved
only for the clump medium, i.e.\ for the enhanced density $\rho_{\rm
C} = D\rho$. From the obtained population numbers, the non-LTE opacity
and emissivity of the clump matter, $\kappa_{\rm C}(D\rho)$ and
$\eta_{\rm C}(D\rho)$, can be calculated.

In the radiative transfer equation, the smooth-wind opacity and
emissivity $\kappa(\rho)$ and $\eta(\rho)$ must be replaced for a
clumped wind by 
\begin{equation}
\kappa_{\rm f} = f_{\rm V}\ \kappa_{\rm C}(D \rho) ~~~\mathrm{and}~~~ 
\eta_{\rm f} = f_{\rm V}\ \eta_{\rm C}(D \rho)\   .
\label{eq:kappa_D}
\end{equation}
The factor \(f_{\rm V}\) accounts for the fact that only this fraction
of a ray actually crosses clumps. {This treatment of radiative
transfer is sometimes termed  the ``filling factor approach'' 
(Hillier \& Millier \cite{HiMi99}).}

The atomic processes contributing to the opacity and emissivity scale
with different powers of the density. For processes {\em linear} in
density, $f_{\rm V}$ and $D$ cancel, but contributions scaling with
the {\em square} of the density (bound-free emission, or free-free
absorption and emission) are effectively enhanced by a factor of
$D$. (Of course, $\kappa$ and $\eta$ are also indirectly affected by
clumping via the population numbers, due to {\nchange modified} rates
in the statistical equations.)

Empirical mass-loss diagnostics are widely based on processes that
scale with the square of the density. The wind emission lines in
Wolf-Rayet and O stars, including H$\alpha$, form in de-excitation
cascades that are fed by radiative recombination. The thermal radio
emission from stellar winds is due to the free-free process.

{When the wind is clumped in the regions from which emission emerges,
the radiation flux is enhanced by a factor of $D$ compared to a
homogeneous model with the same mass-loss rate.  Consequently,} when a
given (radio or line) emission is analyzed with a model that accounts
for microclumping, the derived mass-loss rate will be lower by a
factor of $\sqrt{D}$ than from with a smooth-wind model.

{Puls \etal\ (\cite{Puls_etal2006}) studied the O-type supergiant
$\zeta$\,Puppis and found that the mass-loss rates derived from
H$\alpha$ and from IR/radio continuum emission can only be reconciled
if the wind is strongly clumped ($D \approx 5.5$) already at a radius
of $1.12\,R_\ast$ where the wind velocity is only
$\approx$100\,km\,s$^{-1}$.}

{Line absorption coefficients depend linearly on density. In the case
of resonance lines, the same holds for the re-emitted (``scattered'')
photons. Hence, P Cygni profiles are independent of
microclumping.  Of course, clumping might indirectly affect
resonance lines via the ionization balance. In most cases of practical
importance, the resonance line belongs to the leading ionization stage,
which is robust against such effects.}

\section{{Radiative transfer with clumps of arbitrary optical thickness}}

The basic assumption of the ``microclumping'' approximation described
in the previous section is that the clumps are small compared to the
mean free path of photons, i.e.\ that they are optically thin. Given
the large atomic opacity in the center of typical spectral lines, this
approximation is not generally justified. In the present paper we
therefore relax this approximation in favor of a more general
treatment that accounts for the possibility of clumps being optically
thick at some frequencies. We refer to this generalization as
``macroclumping''.

\subsection{Basic assumptions}

\label{sect:macroclumping.basic}

\begin{figure}[!tb]
\centering
\epsfxsize=\columnwidth
\mbox{\epsffile{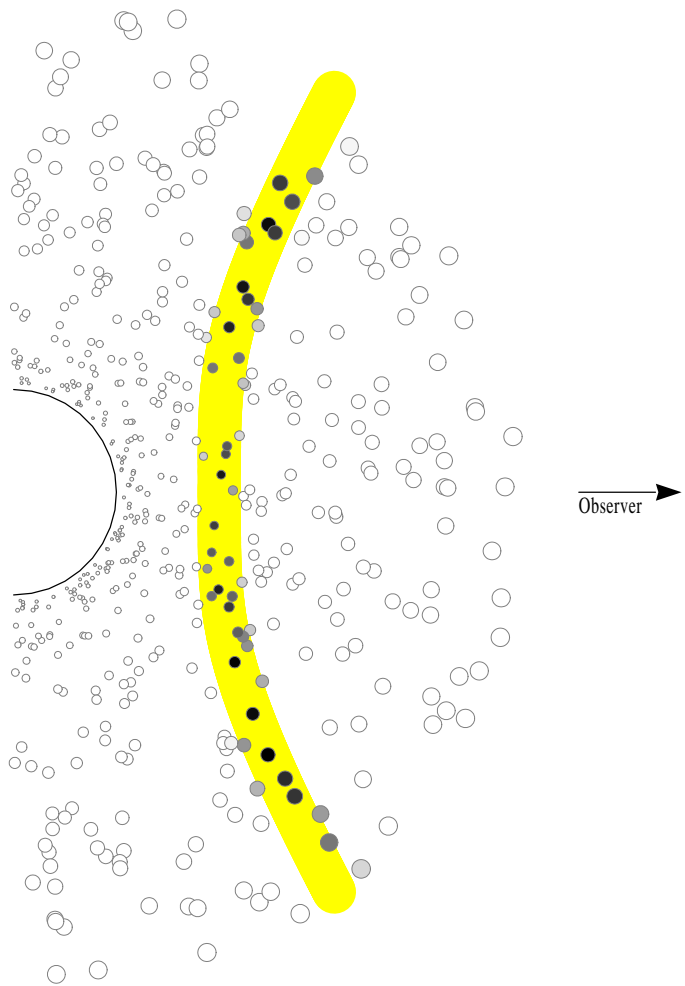}}
\caption{Sketch of a clumped stellar wind. In a smooth wind, rays of a
given observer's frame frequency encounter line opacity only close to
the ``constant radial velocity surface'' (CRVS, thick shaded line). In
a clumpy wind, assuming that the clumps move with the same velocity
law {\nchange as in the homogeneous wind}, only those clumps interact
with the ray that lie close to the corresponding CRVS (dark-shaded
circles). All other clumps are transparent (open circles) if the
continuum opacity is small, so the wind is porous with respect to line
absorption, even when the total volume is densely packed with clumps.}
\label{fig:sketch}
\end{figure}

{To introduce our treatment of porous stellar winds, we first
consider clumps that are smaller than the photon mean free path at any
frequency. Throughout the paper, we refer to these as ``microclumps''.
The non-LTE population numbers that determine opacity and emissivity
are obtained by solving the rate equations for the clump matter
density. The radiation field that enters the radiative transition
rates is calculated with these opacities and emissivities, corrected
for the clump filling factor (cf.\ Eq.\,(1)).

As a next step, consider a few microclumps being merged together to
form one bigger ``macroclump'' without changing the matter
density. The size of this macroclump may now exceed the photon mean
free path at frequencies with large atomic cross sections. In other
words, the macroclump can be optically thick in the cores of a few
strong spectral lines, while it remains optically thin at all other
frequencies.

The non-LTE population numbers depend on the population and
de-population processes, i.e.\ on the collisional and radiative
transition rates.  Because we assume that the clump density does not
change by assembling the macroclump, collisional rates are not
affected and are the same as calculated in the microclump
approximation. The same holds for the radiative rates of optically
thin transitions.

The only transition rates that change in the macroclump are those
radiative rates that are excited at frequencies where the macroclump
is optically thick. The radiation field in optically thick lines may
change drastically in the inner parts of a clump. However, in the
optically thick limit, radiation is trapped inside the clump.  Photons
in these lines are immediately re-absorbed within the clump and do not
escape to the interclump space.

The energy of photons in optically thick lines must be redistributed
elsewhere: the population of the upper line level will increase until
it is de-populated by other (radiative or collisional)
transitions. Hence, in principle the radiation field in any other,
optically thin line may be modified. We neglect this second-order
effect and focus on the macroclumping effect on strong lines.

Photons in optically thick lines that emerge from the clump stem from
its outer layers of optical depth about unity, irradiated from the
interclump space. Therefore, even at these frequencies it is a
reasonable approximation to adopt the radiation field from the
microclumping model to evaluate the radiative transition rates,
i.e. to neglect the feedback of macroclumping effects.

Summarizing these arguments, we find that \\
{\em (i)} most of the transition rates are not affected at all by
macroclumping (because they are collisional or excited by frequencies for
which the clumps are optically thin), and \\
{\em (ii)} for the few optically
thick transitions, the average radiation field neglecting macroclumping 
provides a rough approximation to evaluate the radiative transition rates 
in the outer layers of the clumps. \\
Therefore it is a reasonable first approach to use the non-LTE
population numbers from the microclumping model and to account for the
macroclumping effects only in the ``formal integral''.}

\subsection{{\changed Macro-clumping approach}}

For the radiative transfer in the ``formal integral'', we use a
statistical approach that holds within the limit of large clump number.
As in the microclumping case, we assume that the density inside the
clumps is enhanced by a factor of $D$ compared to a smooth model with
the same mass-loss rate. The interclump medium is void, and the volume
filling factor is $f_{\rm V} = D ^{-1}$.

{So far, no 3-D simulations of stellar winds are available. Hydrodynamic
1-D simulations of wind instabilities by Feldmeier
\etal\ (\cite{Feldmeier_etal1997}) indicate that the clumps take the
form of shell fragments.  This is supported by  detailed modeling
of observed X-ray emission line profiles (Oskinova \etal\
\cite{Oskinova2006}). On the other hand, Dessart \& Owocki (\cite{do05}) 
suggest that clumps may have a similar lateral and radial scale.
However, they emphasize that the clump scale, compression level, and
degree of anisotropy are still uncertain.

To avoid the introduction of an excessive number of free parameters,
we assume here that clumps are isotropic. They have a linear size $l$
that is uniform at a given distance from the stellar center but may be
a function of radius $r$. For simplicity we assume that a clump of
volume $l^3$ has diameter $l$ for any crossing ray, thus neglecting
any center-to-limb variation across the clump.}

It is important to realize that we are considering {\em line}
radiation transfer in a moving medium, which is transparent in the
continuum. The situation is illustrated in Fig.\,\ref{fig:sketch}. A
specific ray, when crossing the stellar wind, only encounters line
opacity over a short part of its path, where the projected expansion
velocity has just the right value to Doppler-shift the line-center
frequency to the selected observer's frame frequency. In the framework
of the Sobolev approximation, the observed line emission or absorption
arises when the ray intersects the corresponding ``constant radial
velocity surface'' (CRVS). In the case of a porous medium, the CRVS
dissolves into patches.  Adopting a clump/void density structure but
an undisturbed velocity field, only those parts of the CRVS are
effectively left that fall into the volume of a clump. Therefore,
porosity effects can be especially important for {\em line} radiation
transfer in expanding atmospheres.

We assume that the clumps are statistically distributed, having an
average separation $L$ between their centers ($L(r)$ varies with
radial location); hence, the volume filling factor is $f_V = l^3 /
L^3$, or
\begin{equation}
D = L^3 / l^3\ . 
\label{eq:dl}
\end{equation}

{\changed 
\noindent From Eq.\,(\ref{eq:dl}) it follows that
\begin{equation}
n_{\rm C} \equiv L^{-3} = D^{-1}\,l^{-3},  
\label{eq:nc}
\end{equation} 
where $n_{\rm C}$ denotes the number density of 
stochastically distributed clumps. }
   
Lucy (\cite{lucy07}) has shown that photospheric turbulence plays a
decisive role in regulating the mass flux of a stationary flow. It is
known from many studies that a high ``microturbulence'' is needed to
reproduce the detailed shape of line profiles  observed in O and
WR star spectra. We identify this ``microturbulence'' $\varv_{\rm D}$
with the typical velocity dispersion within a clump, caused by
stochastic motions and the expansion of the clump itself. In our
radiative transfer calculations, we therefore take the line absorption
coefficient  with a Gaussian profile corresponding to the
Doppler-broadening velocity $\varv_{\rm D}$ and evaluate the optical
depth across one clump formally as for a static medium.

We have already introduced the opacity $\kappa_{\rm C}$ {\em within} a
clump, so we can write the optical depth across one clump with the
help of Eq.\,(\ref{eq:kappa_D}) as
\begin{equation}
\tau_{\rm C} = \kappa_{\rm C}\  l  = 
\kappa_{\rm f}\  D l  \ ,
\label{eq:kdl}
\end{equation}

\noindent Alternatively, $\tau_{\rm C}$ can be expressed by the density 
contrast  $D$ and the average clump separation $L$,
\begin{equation}
\tau_{\rm C} = \kappa_{\rm f}\ D^{2/3}\ L  \ .
\label{eq:tau_C}
\end{equation}
Note that for $L \rightarrow 0$ the clumps become optically thin, 
approaching the microclumping approximation. 

The effective opacity $\kappa_{\rm eff}$ of the clumpy medium is obtained 
in analogy to the usual opacity from atomic absorbers:
\begin{equation}
\kappa_{\rm eff} = n_{\rm C}\ \sigma_{\rm C}\ .
\label{eq:eff}
\end{equation}
Here $\sigma_{\rm C}$ is the effective cross section of a
clump. {\changed ``Effective'' means that the geometrical cross
section, $l^2$, is multiplied by the fraction of photons that is
absorbed when crossing the clump. As was derived in Feldmeier et
al.\ (\cite{Feldmeier2003}),}
 
\begin{equation}          
\sigma_{\rm C} = l^2\  \left( 1 - e^{-\tau_{\rm C}} \right)\ .
\label{eq:sig}
\end{equation}

Using Eq.\,(\ref{eq:nc}) for the number of clumps and 
Eq.\,(\ref{eq:sig}) for the effective cross-section 
in Eq.\,(\ref{eq:eff}),  the effective opacity  becomes 
\begin{equation}
\kappa_{\rm eff} = (D l)^{-1}\ 
\left( 1 - e^{-\tau_{\rm C}} \right)\ .
\label{eq:kapdl}
\end{equation}
In the limit of opaque clumps,   
\begin{equation}
\lim_{\tau_C \rightarrow \infty}  \kappa_{\rm eff} = (D l)^{-1}\ .
\label{eq:h} 
\end{equation}

Noticing from Eq.\,(\ref{eq:kdl}) that 
$ (D l)^{-1} = \kappa_{\rm f} / \tau_{\rm C}$, we obtain a scaling of 
effective opacity with the opacity obtained in the microclumping 
approximation,
\begin{equation}
\kappa_{\rm eff} = 
\kappa_{\rm f}\ \frac{1 - e^{-\tau_{\rm C}}}{\tau_{\rm C}}\,\equiv \,
 \kappa_{\rm f}\ C_{\rm macro}\ .
\label{eq:kef} 
\end{equation}
The factor $C_{\rm macro}$ thus describes how macroclumping changes the
opacity, compared to the microclumping limit. Note that 
the small-clump approximation $\kappa_{\rm eff} \approx \kappa_{\rm f}$ 
is recovered  for optically thin clumps ($\tau_C \ll 1$). For optically 
thick clumps ($\tau_C \msim 1$), however, the effective opacity is 
reduced by a factor $C_{\rm macro}$ compared to the microclumping 
approximation. 

To understand the reduction of opacity in a medium with optically
thick clumps, one can imagine that atomic absorbers are hidden in
optically thick clumps and therefore not contributing to the effective
opacity. Relatedly, Brown \etal\ (\cite{brown04}) noticed that this
also leads to a reduction in radiative acceleration.

{\changed While Feldmeier et al.\ (\cite{Feldmeier2003}) used
Eqs.\,(\ref{eq:eff}) and (\ref{eq:sig}) for their analysis of X-ray
emission line profiles, Eq.\,(\ref{eq:kef}) was derived by Owocki et
al.\ (\cite{ow04}) for the gray opacity. These authors define a
``porosity length'' $h \equiv L^3/l^2$.  Rewriting Eq.\,(\ref{eq:kef})
for the porosity length, one obtains $\kappa_{\rm eff} = h^{-1} \ ({1 -
e^{-\tau_{\rm C}}})$. This can be compared to Eq.\,(A.7) in Feldmeier
\etal\ (\cite{Feldmeier2003}) and Eqs.\,(29) and (31) in Oskinova
\etal\ (\cite{Oskinova2004}), where the opacity is allowed to be
angular-dependent.}

{\changed 
We evaluate the correction factor $C_{\rm macro}$ by specifying the clump
separation $L(r)$, which enters Eq.\,(\ref{eq:tau_C}) in combination with
the density contrast $D(r)$. The reason for this parameterization is that
$D(r)$ must be specified for the density in the non-LTE rate equations.  
Note that the porosity length $h = D^{2/3} L$ alone is not adequate to 
define the model.

Clumps are assumed to be preserved entities. This assumption is
corroborated by the hydrodynamical models (Feldmeier \etal\
\cite{Feldmeier_etal1997}; Runacres \& Owocki \cite{ro02}; Dessart
\& Owocki \cite{do05}), that predict that the 
over-dense structures accelerate radially with $\varv(r)$ and, in
general, do not split or merge.}  As a result, the number density of
clumps has to obey the equation of continuity,
\begin{equation}
n_{\rm C} \propto (r^2\ \varv(r))^{-1}\ .
\end{equation}
Since the average clump separation is $L = n_{\rm C}^{-1/3}$, we can write 
\begin{equation}
L(r) = L_0\  \left(r^2 \varw(r)\right)^{1/3}, 
\label{eq:clump_sep}
\end{equation}
where $L_0$ is the free parameter of our macroclumping model. Note
that $r$ and $L_0$ are in units of the stellar radius, and $\varw(r) =
\varv(r)/\varv_\infty$ is the velocity in units of the terminal speed. 

Plausibly, $L_0$ is close to unity. In the 1-D hydrodynamic
model of Feldmeier et al.\ (\cite{Feldmeier_etal1997}), the wind
shocks evolve from photospheric seed perturbations that have periods
close to the acoustic cutoff. As the latter is close to the wind-flow
time $R_*/\varv_\infty$ for a typical O star, the predicted radial
separation between the dense shells is approximately one stellar
radius.

When studying the effect of macroclumping on the line spectra, it is
important not to spoil the reference continuum. Therefore we only
allow for macroscopic clumping in the line-forming regions, but
suppress it in the photosphere. Photospheric clumping would alter the
radiation field in the line-forming regions (and thus the population
numbers!) and would also change (enhance) the reference continuum for
the line spectrum.

Therefore we augment Eq.\,(\ref{eq:clump_sep}) for the clump separation 
$L$ by a factor $s(\tau_{\rm Ross})$, which switches the macroclumping off at
large Rosseland optical depth, with a linear transition between $\tau_1$ and
$\tau_2$:
\begin{equation}
s(\tau_{\rm Ross}) = \left\{
\begin{array}{lcl}
0 & & {\rm if}~ \tau_{\rm Ross} > \tau_2 \\
1 & & {\rm if}~ \tau_{\rm Ross} < \tau_1 \\
(\tau_2 - \tau_{\rm Ross})/(\tau_2 - \tau_1) & & {\rm else}
\end{array} \right.
\end{equation}
where we usually choose $\tau_1$ = 0.3 and $\tau_2$ = 1.0. We have
checked that the continuum flux (SED) is not modified by macroclumping
effects with these parameters.

We implemented the described modifications of the radiative transfer
into the ``formal integral'' of the PoWR model atmosphere code. At
each integration step along each ray of given frequency, we calculated
the clump optical depth $\tau_{\rm C}$ from Eq.\,(\ref{eq:tau_C}) and
evaluated the macroclumping correction factor (Eq.\,(\ref{eq:kef})),
which is applied to the opacity before the integration step is
performed. Note that the same correction factor must also be applied
to the emissivity to preserve the non-LTE source function according to
our assumption that the population numbers are not affected.

\subsection{Clump separation and the total number of clumps}
\label{sect:Nc}

{The equations in the previous section describe our approach to
optically thick clumping. Only one additional free parameter, $L_0$,
has been introduced compared to the microclumping approach
(Eq.\,\ref{eq:clump_sep}).  For a specified stellar atmosphere model,
the choice of $L_0$ determines to what extend the clumps are
optically thick at line frequencies. Within the limit of small $L_0$, 
the macroclumping effects become negligible.  Realistic values of $L_0$
can be estimated from considering the total number of clumps that
reside in the wind at any given moment.}

The number of clumps per unit volume, i.e.\ their number density,
{\nchange which is a statistical variable}, follows from their average 
separation 
\begin{equation}
n_{\rm C}(r) = L^{-3} = \left( L_0^3\ r^2\ \varw(r)  \right)^{-1}\ .
\end{equation}

\noindent Thus the total number of clumps that are found within the 
radial range from $r_1$ to $r_2$ is

\begin{equation}
N_{\rm C} = \int_{r_1}^{r_2} n_{\rm C}(r)\ 4\pi\ r^2\ {\rm d}r = 
\frac{4\pi}{L_0^3} \int_{r_1}^{r_2} \frac{{\rm d}r}{\varw(r)}
= \frac{4\pi}{L_0^3} (t_2 - t_1)\ ,
\label{eq:NC}
\end{equation}
where $t_2 - t_1$ is the flight time between $r_1$ and $r_2$ in 
units of the dynamical time scale, $R_*/\varv_\infty$. For ``beta 
velocity laws'' $\varv(r) = \varv_\infty (1 - 1/r)^\beta$, 
analytical solutions for certain values of $\beta$ are given 
in Hamann et al.\ (\cite{DAC-paper}). For $\beta = 1$ it holds that 
\begin{equation}
t(r) = \int \frac{{\rm d}r}{\varw(r)} = r + \ln (r-1) + \mathrm{const}\ .
\end{equation}

Let us consider, for instance, the range between $r_1$ = 1.05, 
where the wind velocity is 5\% of the terminal speed, and $r_2$ = 10 
stellar radii. The total number of clumps in this volume obtained 
from the previous two equations is 
\begin{equation}
N_{\rm C} = 178\ L_0^{-3}\ .
\label{eq:NC-10}
\end{equation}
Note that $N_{\rm C}$  scales with $r_2$ more weakly than linearly;  
when restricting the volume to $r_2$ = 5, the number of enclosed 
clumps is still $N_{\rm C} = 105\ L_0^{-3}$.

{\changed Empirically, not much is known about the number of clumps in
stellar winds. {\nchange So far, only rough order-of-magnitude
estimates can be made.}  Line profile variability may provide some
relevant information. L\'epine \& Moffat (\cite{Lepine+Moffat1999})
monitored nine WR stars and found stochastic line profile variations
in the form of narrow ($\approx 100$\,km/s) emission sub-peaks on top
of emission lines. They explained their data by $10^3 \lsim N_{\rm
C}\lsim 10^4$ ``blobs'' being present at any time in the line emission
region. Using Eq.\,(\ref{eq:NC-10}) this corresponds to $L_0 \approx
0.3$.  We will see in the next sections that a value of $L_0$ between
0.2 and 0.5, implying a plausible number of $10^3$ to $10^4$ clumps in
the line-forming region, is high enough to produce significant
macroclumping effects on the emergent spectra.

Further support for selecting $L_0$ in the range 0.1 to 1.0 comes from
analizing X-ray emission lines. In Oskinova et al.\
(\cite{Oskinova2006}), we reproduced the {\em Chandra} observations of
O-star line profiles by assuming that spherical shells are ejected
once per dynamic time scale $t_{\rm fl}\,=\,R_*/\varv_\infty$, which
corresponds to a radial separation of one stellar radius when
$\varv_\infty$ is reached.  Note, however, that there are subtle
differences between the ``fragmented shell model'' used for the X-ray
modeling and the present paper, thereby preventing a direct comparison
of the parameters. }

\section{The impact of macroclumping on O star spectra: 
$\zeta$\,Puppis }

\begin{figure}
\centering
\epsfxsize=\columnwidth
\mbox{\epsffile{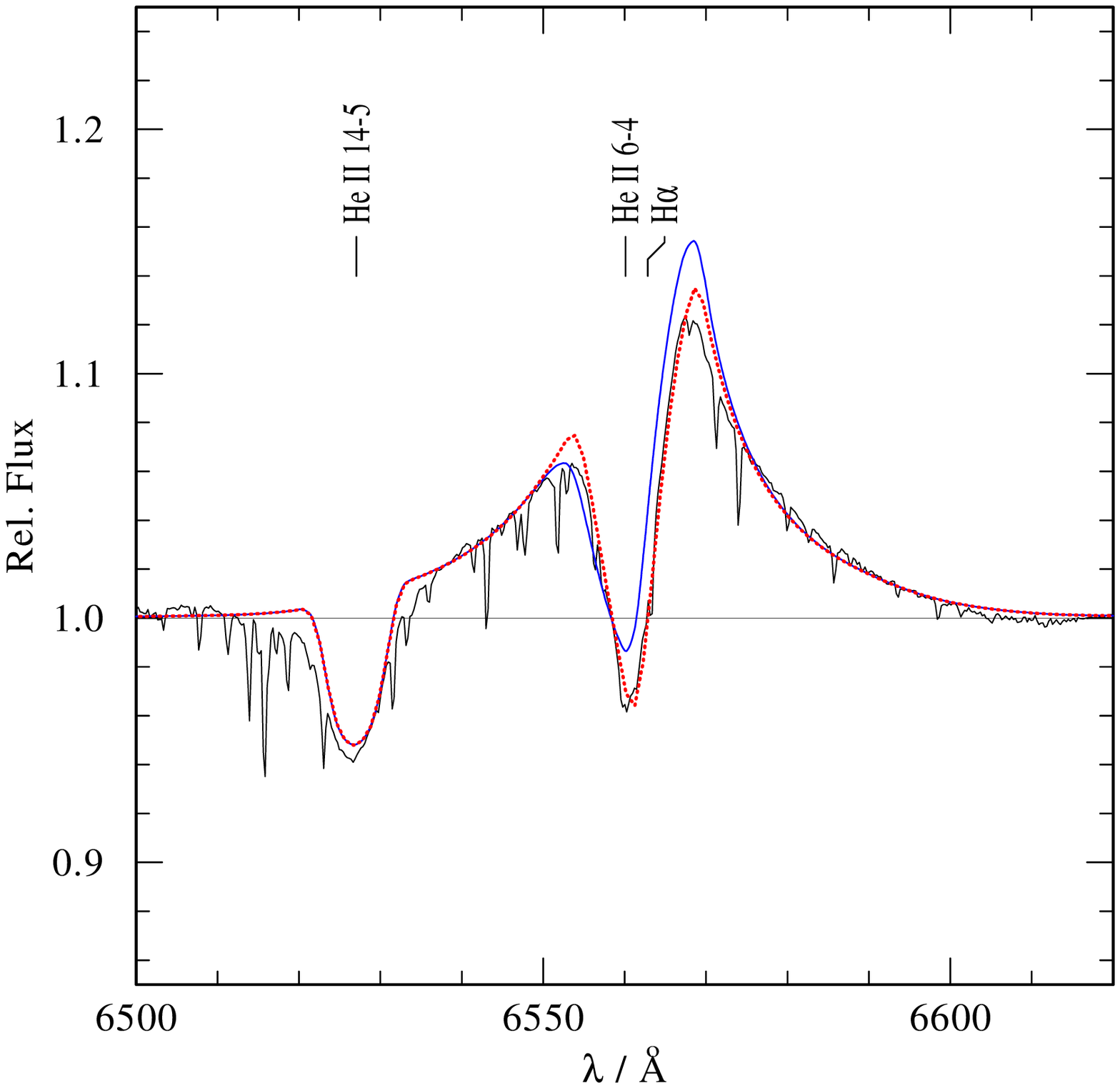}}
\caption{
Effect of macroclumping on the H$\alpha$ line from an O star. The
observed spectrum of $\zeta$\,Pup is shown for comparison (black, ragged
line). The model parameters (see text) correspond to $\zeta$\,Pup. 
The solid (blue) line results from a model that
only accounts for small-scale clumping with a density contrast $D$ = 10,
switched on at the sonic point (10\,km/s). The dotted (red) curve,
partly identical with the solid (blue) curve, is obtained with 
macroclumping in the formal integral (also switched on at the sonic
point with clump-separation parameter $L_0$ = 0.2). The effect of 
macroclumping on the H$\alpha$ emission is small.}
\label{fig:compare-zPup-Halpha}
\end{figure}

As a representative test, we demonstrate the influence of
macroclumping on the spectrum of the O-type supergiant
$\zeta$\,Puppis. First we calculate a non-LTE atmosphere model with
the PoWR code (see Hamann \etal\ \cite{PoWR} for details). Based on
the model's population numbers, the emergent spectrum is then
obtained by solving the ``formal integral''. This step is done
twice, once with the usual code version that accounts for clumping
only in the small-scale limit, and a second time with the
macroclumping formalism as developed in the previous section.

\begin{figure*}[!bt]
\centering
\epsfxsize=0.8\textwidth
\mbox{\epsffile{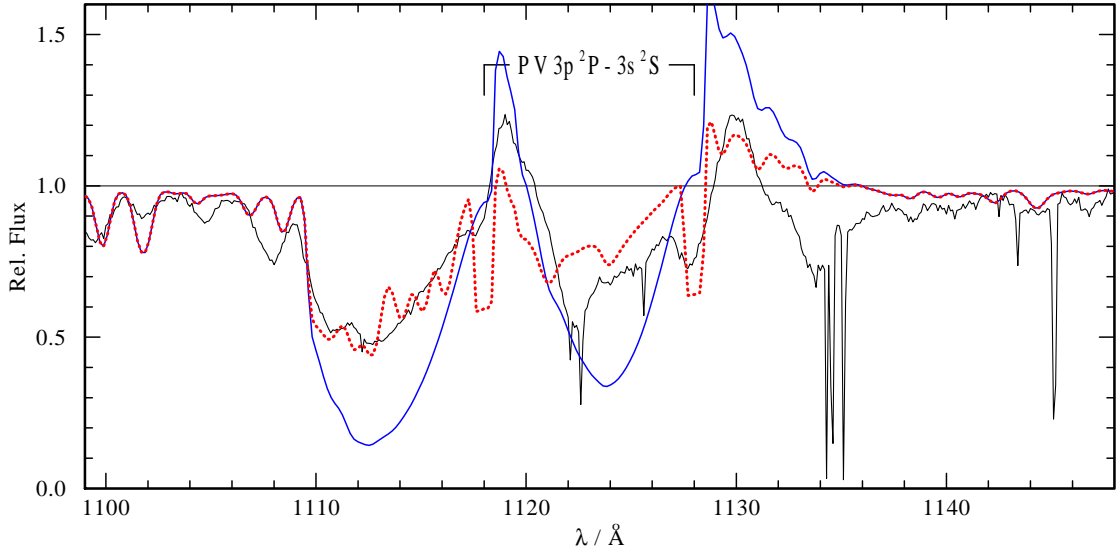}}
\caption{Effect of macroclumping on the P\,{\sc v} resonance doublet at
1118/1128\,\AA\ for an O star. The spectrum of $\zeta$\,Pup as
observed by {\sc copernicus} is shown for comparison (black, ragged
line).  The model parameters (see text) correspond to
$\zeta$\,Puppis. We adopt a microturbulence velocity of 50\,km/s in
the formal integral and do not apply rotational broadening; therefore
the photospheric absorption components are not properly reproduced.
Numerous weak spectral features in this range are due to iron.  The
usual microclumping modeling yields P\,Cygni features that are too
strong (blue, continuous line). With our macroclumping formalism, the
line features are reduced to the observed strength (red, dotted
curve). Macro-clumping was switched on at about the sonic point
(10\,km/s) with a clump-separation parameter $L_0$ = 0.2. }
\label{fig:PV}
\end{figure*}

For the macroclumping simulation, we set the clump separation parameter
to $L_0 = 0.2$.  According to Eq.\,(\ref{eq:NC-10}), this implies a
$2.2 \cdot 10^4$ clumps in the line-forming region (within 10\,$R_*$).
This number of clumps is large enough to justify our statistical
treatment, and it compares well with the empirical estimates (cf.\
Sect.\,\ref{sect:Nc}).

Macroclumping is assumed to start at about the sonic point (10\,km/s),
together with microclumping.
{As model parameters for $\zeta$\,Puppis, we adopt (compare Puls et al.\
\cite{Puls_etal2006}) $\log L/L_\odot$ = 5.9, $T_*$ = 39\,kK, $\log
\dot{M}/(M_\odot {\rm yr}^{-1})$ = $-5.6$. We use the commonly
accepted $\beta$-law for the velocity field, but split it into two
ranges with $\beta$=0.9 from the photosphere to 0.6\,$\varv_\infty$
and $\beta$=4 for the outer part of the wind. The terminal speed
$\varv_\infty$ is 2250\,km/s. The density enhancement factor is set to
$D$ = 10 in the supersonic part of the atmosphere. For the chemical
composition, we adopt 42/56/0.12/0.4/0.9/0.14 (mass fractions in \%)
for the H/He/C/N/O/Fe-group (Repolust \etal\ 2004). We also included
phosphorus with solar abundance ($6.15\times 10^{-6}$ mass fraction).}

{We concentrate here on two lines, H$\alpha$ and the P\,{\sc v}
resonance doublet. The H$\alpha$ emission is the most important
diagnostic tool for O-star mass-loss rates. The P\,{\sc v} doublet,
unlike other resonance lines, is not saturated (because of the low
phosphorus abundance) and therefore potentially useful for the
empirical determination of mass-loss rates. FUSE observations of the
P\,{\sc v} resonance doublet in the extreme ultraviolet caused the
recent debate on a drastic downward revision of mass-loss rates
(Fullerton et al.\ \cite{Fullerton06}, 
Bouret et al.\ \cite{Bouret_etal2005}).}

{The H$\alpha$ emission line is strongly affected by microclumping,
especially in the lower regions of the wind (see also Puls \etal\
\cite{Puls_etal2006}). 
In contrast to H$\alpha$, resonance lines are not sensitive to
microclumping because of their linear dependence on density. 
Moreover, P\,{\sc v} is the leading ionization stage of phosphorus in
the wind of typical O stars. This makes the line robust against details
of the modeling, including the ``superionization'' by X-rays, which is
an essential effect for the O\,{\sc vi} and N\,{\sc v} resonance lines.
For the same reason, the P\,{\sc v} resonance doublet is also
insensitive to the second-order effects on the population numbers that
we neglect in our macroclumping approach, as discussed in
Sect.\,\ref{sect:macroclumping.basic}.

Figure \ref{fig:compare-zPup-Halpha} displays the synthetic spectra
from both model versions, without macroclumping use and with our
macroclumping treatment, but otherwise identical parameters. The
observation of $\zeta$\,Pup is shown for comparison. The mass-loss
rate of the model is tuned to the adopted density contrast $D=10$ in
order to obtain a good fit. We find that a model with the same
mass-loss rate, but without microclumping, fails to reproduce the
observed H$\alpha$ profile by far (the line appears in absorption).

Interestingly, the macroclumping effect is not significant for the
H$\alpha$ line. We tested different $L_0$ and found no sensitivity for
any realistic values. The obvious reason is that H$\alpha$ is
optically thin in the supersonic part of the wind from
$\zeta$\,Pup, so accounting for macroclumping does not lead to a
different empirical mass-loss rate from H$\alpha$ fitting.

\begin{figure}[!tb]
\centering
\epsfxsize=\columnwidth
\mbox{\epsffile{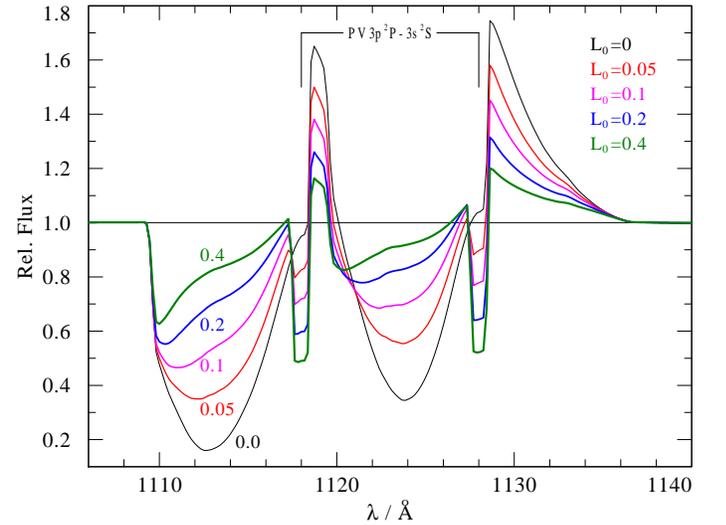}}
\caption{Synthetic P\,{\sc v} resonance doublet at
1118/1128\,\AA\ {\nchange for different values of} the clump
separation parameter $L_0$ (labels). The model parameters are for
$\zeta$\,Pup (see Sect.\,4), with $\varv_{\rm D}=50$ \,km/s.  A clump
separation of zero recovers the microclumping limit.  The larger the
clump separation, the weaker the line becomes.}
\label{fig:Lser}
\end{figure}

When Bouret \etal\ (\cite{Bouret_etal2005}) and Fullerton \etal\
(\cite{Fullerton06}) recently analyzed the P\,{\sc v} resonance
doublet in O-star spectra, they encountered severe discrepancies in
the mass-loss rates inferred from H$\alpha$. They suggested that the
only way to reconcile these different diagnostics is to assume much
higher clumping contrast than assumed hitherto, implying a drastic
reduction of the empirical mass-loss rates.

\begin{figure}[!tb]
\centering
\epsfxsize=\columnwidth
\mbox{\epsffile{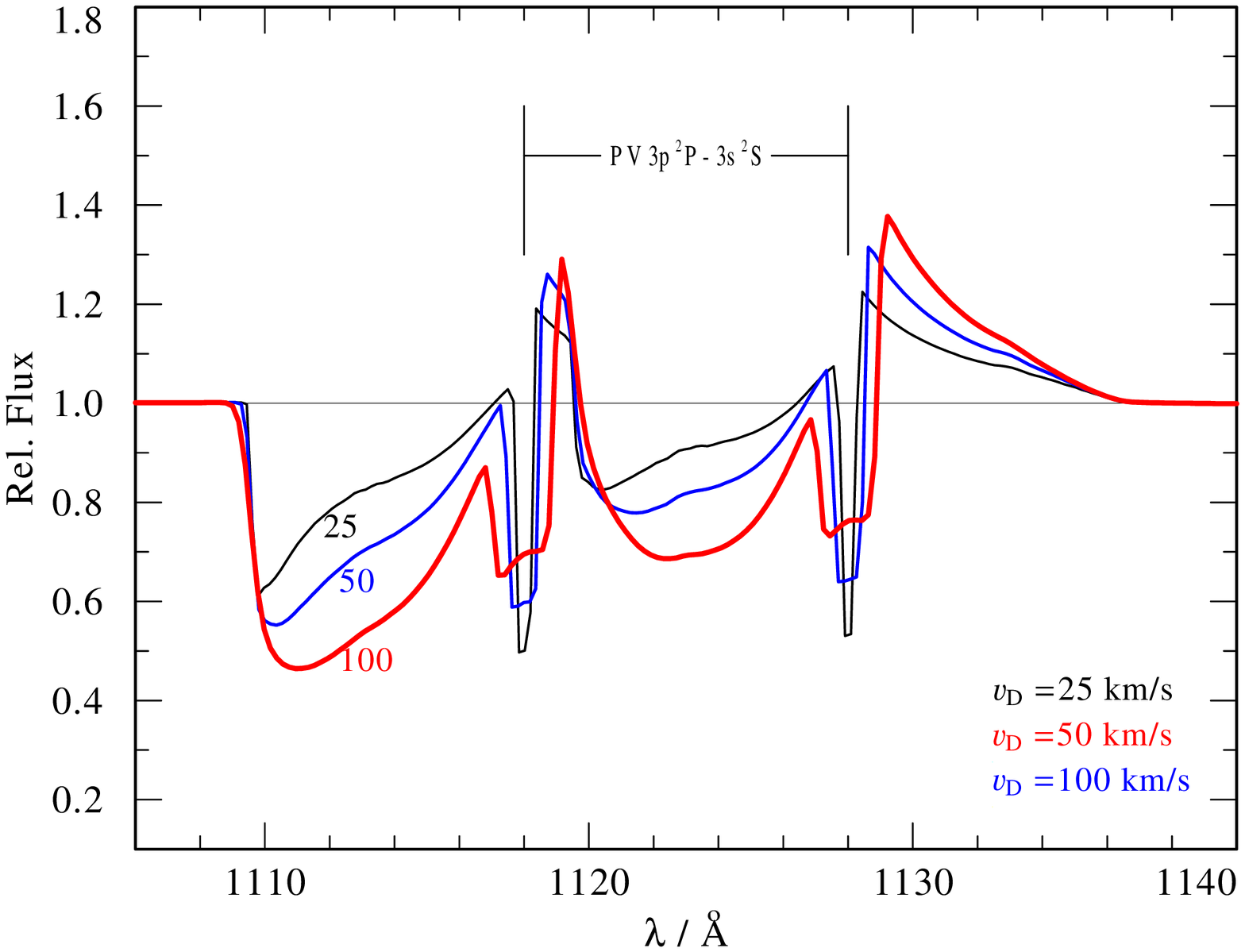}}
\caption{Synthetic P\,{\sc v} resonance doublet at
1118/1128\,\AA\ {\nchange for different values of} the microturbulence 
velocity $\varv_{\rm D}$ (labels). The clump separation parameter is
$L_0\,=\,0.2$ for all cases. {\nchange All other model parameters are
kept fixed.}
When the velocity dispersion across the clumps is decreased, 
the macroclumping effect becomes more pronounced and leads to
a weaker line profile. Note that the unrealistically strong
photospheric absorption features become weaker with lower
microturbulence velocities, as expected from the curve-of-growth 
effect.}
\label{fig:vdser}
\end{figure}

This discordance is illustrated in Fig.\,\ref{fig:PV}, showing the EUV
spectrum of $\zeta$\,Pup (from the {\sc copernicus} satellite) 
with the P\,{\sc v} doublet. The continuous line is
the synthetic spectrum of the conventional model without macroclumping.
The parameters are the same as in Fig.\,\ref{fig:compare-zPup-Halpha},
i.e.\ perfectly able to reproduce the observed H$\alpha$ emission. As
can be seen, the predicted P-Cygni profiles of the P\,{\sc v} doublet
are much stronger than observed. To fit the H$\alpha$ and
P\,{\sc v} lines consistently with the microclumping model, a higher
density contrast $D$ and  a lower mass-loss rate would be required.  
\begin{figure*}[!tb]
\centering
\epsfxsize=0.8\textwidth
\mbox{\epsffile{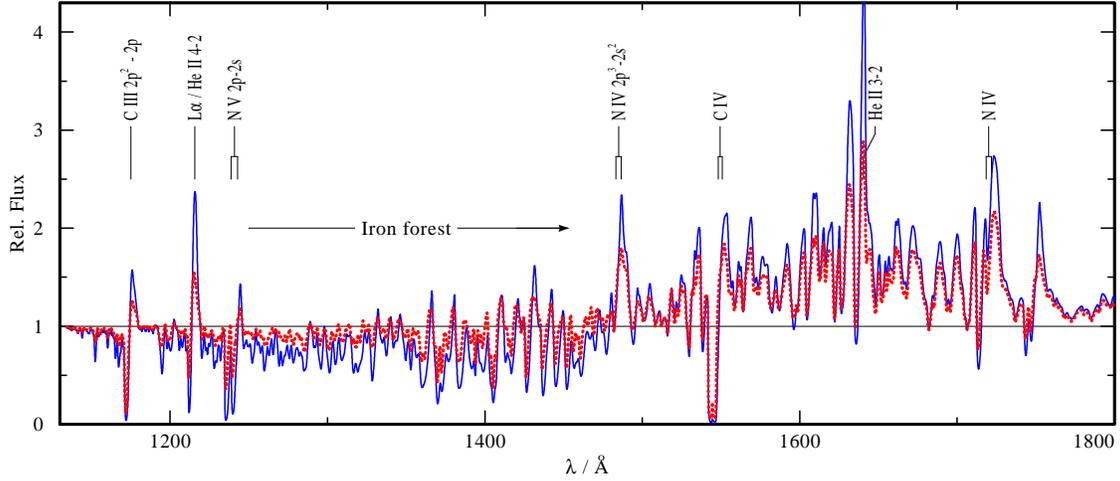}}
\caption{Effect of macroclumping on the short-UV spectrum from a WN star
of late subtype; the model parameters (see text) correspond to WR\,78 (WN7h).
The solid line is calculated with microclumping alone, and the
dotted line is obtained with macroclumping in the formal integral
(clump-separation parameter $L_0$ = 0.5).}
\label{fig:compare-WNL-UV}
\end{figure*}

Adopting the same model, we computed the stellar spectrum with our
macroclumping formalism (Fig.\,\ref{fig:PV}). In the
synthetic spectrum, the P\,{\sc v} features are now drastically
reduced, just to about the observed strength. Note that the numerous
weak iron lines in this spectral range are not affected by the
macroclumping correction. When accounting for macroclumping, the
H$\alpha$ and P\,{\sc v} lines in the observed spectrum of
$\zeta$\,Pup can be fitted simultaneously, using a ``standard''
clumping factor of $D$\,=\,10 (filling factor $f_V$\,=\,0.1). We
conclude that accounting for macroclumping {\changed can} result in a
concordance of mass-loss estimates.

\begin{figure*}[!tb]
\centering
\epsfxsize=0.8\textwidth
\mbox{\epsffile{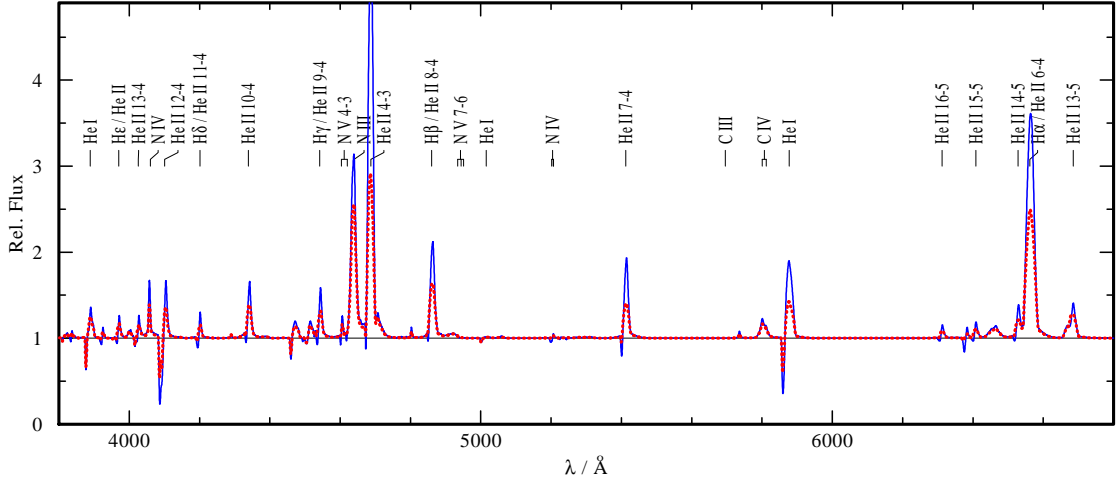}}
\caption{Same as Fig.\,\ref{fig:compare-WNL-UV}, but for the 
optical range} 
\label{fig:compare-WNL-OPT}
\end{figure*}

\section{\changed Influence of clump separation parameter, $L_0$, and 
velocity dispersion within a clump, $\varv_{\rm D}$ }
\label{sect:dif}

{\changed As explained in Sect.\,3.2, $L_0$ is the essential free
parameter of our macroclumping model. To demonstrate its influence, we
consider the P\,{\sc v} resonance doublet, keeping all other
parameters as in the previous section. Blending iron lines are
suppressed now for clarity.

We start by computing a model without macroclumping correction, i.e.\
$C_{\rm macro}=1$ implying $L_0=0$ from Eq.\,(\ref{eq:tau_C}). The
resulting profile is shown in Fig.\,\ref{fig:Lser}.
As the next step, we increase $L_0$ stepwise while all other parameters
are kept constant. Increasing $L_0$ leads to an increased clump
optical depth $\tau_{\rm C}$. Therefore, the macroclumping correction
factor $C_{\rm macro}$ decreases, leading to a reduction of the
effective opacity $\kappa_{\rm eff}$.  As can be seen in
Fig.\,\ref{fig:Lser}, higher values of $L_0$ result in weaker lines.

Significantly, {\nchange the macroclumping effect depends not only on}
the number of clumps (defined by $L_0$), but also on the velocity
dispersion within the individual clumps. This is a specific
consequence of the line radiative transfer in an expanding wind.
Figure\,\ref{fig:vdser} shows model lines that are computed with a
constant parameter $L_0=0.2$, but for different velocities $\varv_{\rm
D}$. As can be seen, the porosity effect of reduced line strengths is
stronger for lower values $\varv_{\rm D}$.

\begin{figure*}[!tb]
\centering
\epsfxsize=0.8\textwidth
\mbox{\epsffile{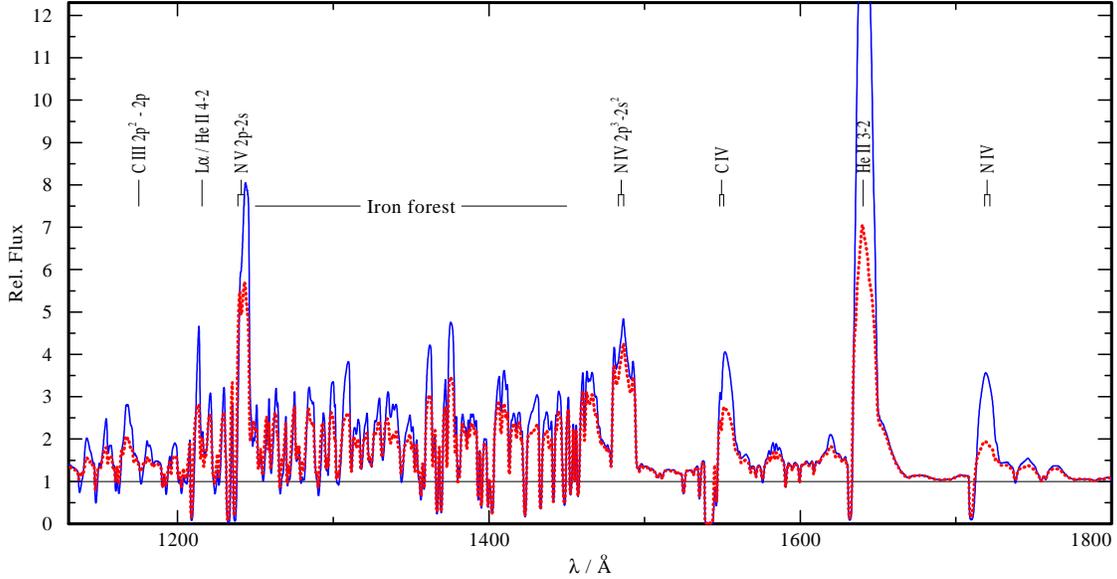}}
\caption{Effect of macroclumping on the UV spectrum from a WN star of
early subtype WR\,7 (WN4-s) with parameters as given in the text. The
solid line is calculated with microclumping alone, and the dotted
line is obtained with macroclumping in the formal integral
(clump-separation parameter $L_0$ = 0.5).}
\label{fig:compare-WNE-UV}
\end{figure*}

\begin{figure*}[!tb]
\centering
\epsfxsize=0.8\textwidth
\mbox{\epsffile{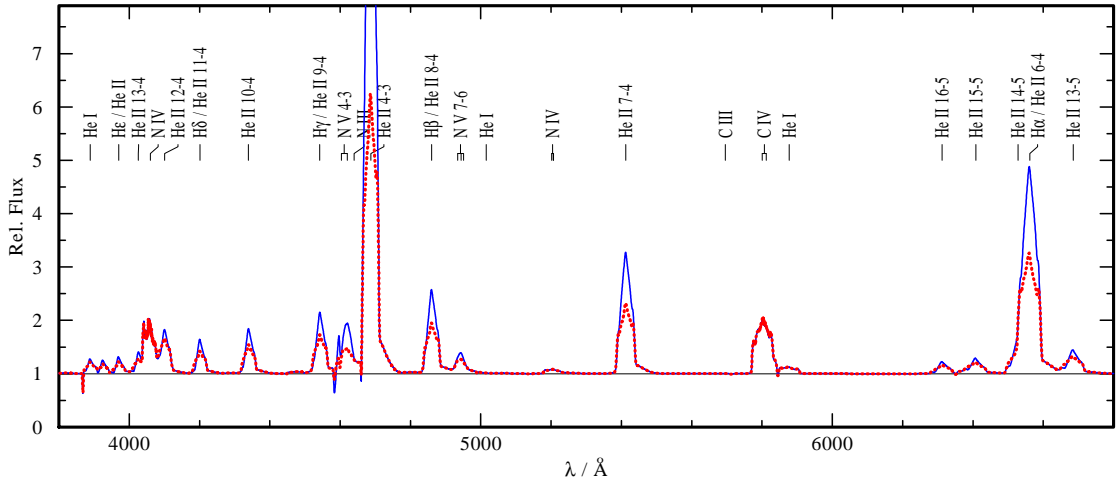}}
\caption{Same as Fig.\,\ref{fig:compare-WNE-UV}, but for the 
optical range.} 
\label{fig:compare-WNE-OPT}
\end{figure*}

To understand this result, we considered the line absorption coefficient in
a clump.  With lower velocity dispersion within a clump, the Doppler
broadening is smaller, and the line absorption profile is narrower
{\nchange but peaks higher}. Thus, the clump optical depth in the
line core becomes larger, while it is smaller in the line wings.
Higher optical depth in the line core leads to a reduction of the
effective opacity and a weakening of the line (see Eq.\,(\ref{eq:kef})).

This is in accordance with our interpretation of a porous constant
velocity surface (see Sect.\,3.2).  As highlighted in
Fig.\,\ref{fig:sketch}, only clumps in the vicinity of the constant
radial velocity surface contribute to the optical depth along the
line-of-sight at a given frequency (see Sobolev \cite{sob}). In a wind
with a monotonic velocity field $\varv(r)$, the characteristic width
of this vicinity is given by the ``Sobolev length'', $\varv_{\rm
D}/(d\varv/dr)$ (e.g. Grinin \cite{gr01}).  For lower
Doppler-broadening velocity $\varv_{\rm D}$, the clump is optically
thick for a narrower range of frequencies within the line.
Consequently the number of clumps that can contribute to the
line-of-sight optical depth is smaller, and the effect of porosity is
greater. 

The velocity dispersion across a clump is well-constrained both from
empirical and theoretical considerations. It has been known for
decades (e.g.\ Hamann \cite{wrh80}, Lamers \etal\ \cite{sei}) that
detailed UV line fits for O stars require a ``microturbulence
velocity'' of 50 to 100\,km/s. The same holds for the emission line
spectra of Wolf-Rayet stars.  In our picture of a clumped wind, this
small-scale Doppler broadening is identified with the typical velocity
dispersion within one clump.  Moreover, it has been noted repeatedly
from monitoring line-profile variability that narrow features that
appear transiently are never narrower than 50\, to \,100\,km/s
(e.g. L\'epine \& Moffat\ \cite{Lepine+Moffat1999}).

Clumps can only survive over the dynamic flow time when their internal
velocity dispersion is small compared to the wind velocity,
$\varv_{\rm D} \ll \varv_\infty$. Close inspection of the 1-D
time-dependent model by Feldmeier \etal\ (\cite{Feldmeier_etal1997})
reveals that, within the density enhancements, the velocity does not
differ by more than $\approx$100\,km/s.  Since $\varv_{\rm D}$ is
constrained to 50 to 100\,km/s for O and WR stars, the clump
separation $L_0$ is the basic free parameter of the macroclumping
model.  }
\section{{Effect of macroclumping on Wolf-Rayet spectra}}
\label{sect:WNLstar} 

{In this section we test the effect of macroclumping on the
emergent spectra for representative models of a WNL star (i.e.\ a
late-subtype Wolf-Rayet star of the nitrogen sequence), and a WNE
(i.e.\ early-subtype Wolf-Rayet) star with strong emission lines.

The two examples} are selected from the {\nchange PoWR} grid of models
for WN stars.  First we take a model with 20\% hydrogen (``WNL grid''
model WNL07-11, Hamann et al. \cite{PoWR}) with parameters
corresponding to a late subtype (WN7): $T_*$ = 50\,kK, $\log R_{\rm
t}/R_\odot$ = 1.0 (see Hamann et al. \cite{PoWR} for the definition
and role of this ``transformed radius''). We use the standard
$\beta$-law for the velocity field with $\beta=1$ and $\varv_\infty$ =
1000\,km/s. The microturbulence velocity for Wolf-Rayet models is set
to $\varv_{\rm D}$\,=\,100\,km/s.  In contrast to the published grid
({\nchange www.astro.physik.uni-potsdam.de/$\sim$wrh/PoWR/powrgrid1.html}),
we choose here a density contrast of $D$ = 10. The adopted composition
is 78/20/1.5/0.01/0.14 for the mass fractions (in
\%) of the He/H/N/C/Fe-group. Except for the different (but scalable)
luminosity, these parameters correspond to the Galactic Wolf-Rayet
star WR\,78. Compared to the previous example ($\zeta$\,Pup), this
model is slightly hotter and has a denser wind. In this example we
have chosen a clump separation parameter of $L_0 = 0.5$, i.e.\ the
total number of clumps within 10\,$R_*$ according to
Eq.\,(\ref{eq:NC-10}) is 1424.

The UV spectrum (Fig.\,\ref{fig:compare-WNL-UV}) shows numerous iron
lines, partly in absorption and partly in emission at longer wavelengths.
There are also a number of strong emission features, mostly
as part of the P\,Cygni profiles. The highest emission peak is due to the
He\,{\sc ii} 1640\,\AA\ transition.  Comparing again the emergent
spectrum with and without the macroclumping effect, we notice that
generally all features, emission and absorption, are reduced in
strength. Only the weakest lines are not affected, whereas the
strongest emissions are cut down by more than 50\%.

In the optical range similar reductions are found for strong lines,
but Fig.\,\ref{fig:compare-WNL-OPT} also demonstrates that the effect
differs between individual lines of similar flux.  {Obviously, to
restore a spectral fit with the macroclumping model, one would have to
increase the mass-loss rate and possibly to re-adjust other parameters
and abundances. }

{Our example with the densest wind corresponds to an early-subtype WN
star with strong lines, WR\,7 (WN4-s).  The parameters are $\log
L/L_\odot$ = 5.3, $T_*$ = 112\,kK, $\log R_{\rm t}/R_\odot$ = 0.3, no
hydrogen, $\varv_\infty$ = 1600\,km/s, $D$=10. The chemical
composition is the same as for the WNL model from
Sect.\,\ref{sect:WNLstar}, but without hydrogen.  Except for the
higher value of $D$, this model corresponds to the PoWR grid model
WNE14-18 (Hamann et al.\ \cite{PoWR}). For the macroclumping
correction, we again choose a clump separation of $L_0 = 0.5$.

The emergent spectra including macroclumping effects are shown in
(Figs.\,\ref{fig:compare-WNE-UV} and \ref{fig:compare-WNE-OPT}).  As
expected, the macroclumping reduction is most pronounced for the
strongest emission lines.  As in the O star example, it is apparent
that higher mass-loss rates must be chosen in order to maintain the
lines as strong as they have been without macroclumping correction.

{\changed Interestingly, even the ``iron forest'' is now significantly
affected by the porosity effect. This is because many of the iron
lines in dense WR winds are quite strong (peak intensities up to four
times the continuum, cf.  Fig.\,\ref{fig:compare-WNL-OPT}), and form
throughout the wind like any other lines of comparable strength.}  }

{\changed 
Figure\,\ref{fig:wnser} demonstrates the influence of the clump
separation parameter $L_0$ on the prominent He\,{\sc ii} emission line
at 4686\,\AA.  The emission decreases with increasing $L_0$. This
porosity effect is already significant for relatively small clump
separation, like  $L_0 = 0.1$.

In the process of a spectral analysis, one would have to compensate
for this macroclumping effect by adopting a higher mass-loss
rate. Changing $\dot{M}$, however, shifts the ionization
balances. Given the different sensitivities of individual lines on the
macroclumping effect (cf.\ Figs.\,6-9), element abundances may become
involved as well. It should also be noticed that the density contrast
$D(r)$ may actually vary with radius. Thus it is by no means
straightforward to achieve a new consistent spectral fit for WR stars
with macroclumping, so we must leave this task for future work.}

\section{Summary}

{We have incorporated the effect of wind porosity into the ``formal
integral'' of the PoWR non-LTE atmosphere code. In a statistical
treatment, optically thick clumps lead to an effective reduction of
the opacity. The feedback of macroclumping on the population numbers
is neglected as a second-order effect, especially in the case of
resonance lines from a leading ionization stage.

The effect of porosity on line formation differs drastically from that
on the medium with continuous opacity. The effect of porosity on lines
is enhanced due to the small extent of the zone where a ray is in
resonance with the line frequency (``scattering zone'', ``constant
radial velocity surface''), while outside this zone the line opacity
vanishes anyway due to the Doppler shift. Although a large
number of clumps may be present in the wind, only a limited number of
these clumps contribute to the line opacity along a specific ray. The
statistical treatment of porosity is corroborated by the
time-averaging nature of observations. Even when the number of clumps
that are present in the scattering zone at a given moment of time is
small, there is a large number of clumps passing through this zone
during a typical integration time of the observation.}

{Following the predictions of time-dependent hydrodynamic models, we
assume that clumps are preserved entities and that there is no
velocity gradient inside the clumps. The velocity dispersion within
individual clumps is described by a ``microturbulence'' velocity that
broadens the line opacity profile. For simplicity, the clumps are
assumed to be isotropic.

\begin{figure}[!tb]
\centering
\epsfxsize=0.8\columnwidth
\mbox{\epsffile{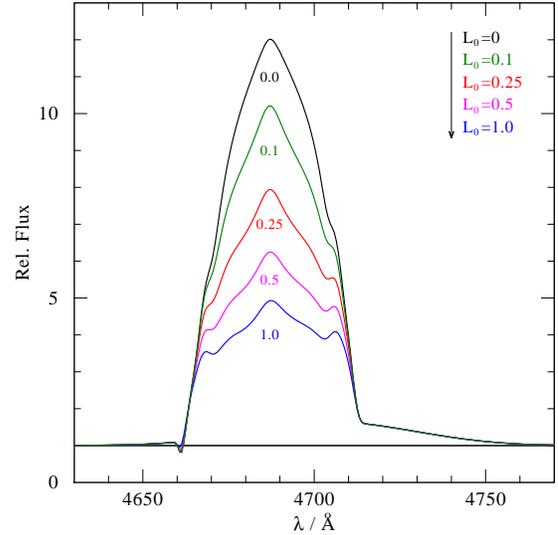}}
\caption{Synthetic  He\,{\sc ii} line at $\lambda$4686\AA\
in dependence on the clump separation parameter $L_0$ (see labels). 
$\varv_{\rm D}=50$ \,km/s for all lines.  A clump separation of
zero recovers the microclumping limit.  The larger is the clump
separation, the weaker the line becomes. }
\label{fig:wnser}
\end{figure}

We modeled the emergent spectrum of the O-type supergiant
$\zeta$\,Puppis and demonstrated the influence of macroclumping on the
mass-loss diagnostic lines H$\alpha$ and P\,{\sc v}. Whereas the
optically thin H$\alpha$ line is not affected by wind porosity, the
P\,{\sc v} resonance doublet becomes significantly weaker when
macroclumping is accounted for. Hence, neglecting macroclumping can
lead to underestimating  empirical mass-loss rates.  In the case
of $\zeta$\,Pup, the discrepancy between the H$\alpha$ and P\,{\sc v}
diagnostic found by Fullerton et al.\ (\cite{Fullerton06}) can be
entirely resolved with the macroclumping modeling, without a downward
revision of the mass-loss rate.

We also studied the influence of macroclumping on the emergent spectra
for two re\-presentative Wolf-Rayet star models, corresponding to a WN
star of late subtype and a WN star of early subtype with strong
lines. The results show that basically all spectral features, in
both emission and absorption, become significantly weaker when
macroclumping is taken into account. Strong lines are typically
reduced by a factor of two in intensity, while weak lines remain
unchanged. }

{\changed In spectral analysis, this weakening of lines must be
compensated for by adopting a {\em higher} mass-loss rate.  For
emission lines that scale with $\rho^2$ because they are fed by
recombination, macroclumping is counteracting microclumping in its
effect on the empirical mass-loss rates.

Most important, wind porosity affects the P-Cygni profiles from
resonance lines, in contrast to microclumping, which does not influence
opacities that depend linearly on density. This can resolve the
reported discrepancies between resonance-line and recombination-line
diagnostics. Future work is needed to quantify this effect for
individual stars.}

\acknowledgements 

This work has been supported by the Deutsche Forschungsgemeinschaft
with grant Fe\,573/3. The authors thank the referee for the thorough
and thoughtful review that led to the significant improvement in the
paper.

\end{document}